# Optical Excitation and Decay Dynamics of Ytterbium Atoms Embedded in a Solid Neon Matrix

C.-Y. Xu[1,2] (徐晨昱), S.-M. Hu[1,3] (胡水明), J. Singh[1], K. Bailey[1], Z.-T. Lu[1,2] (卢征天), P. Mueller[1], T. P. O'Connor[1], and U. Welp[4]

[1]*Physics Division, Argonne National Laboratory, Argonne, Illinois 60439, USA*

[2]*Department of Physics and Enrico Fermi Institute, University of Chicago, Chicago, Illinois 60637, USA*

[3]*Hefei National Laboratory for Physical Science at Microscale, University of Science and Technology of China, Hefei, Anhui 230026, China*

[4]*Material Science Division, Argonne National Laboratory, Argonne, Illinois 60439, USA*

Neutral ytterbium atoms embedded in solid neon qualitatively retain the structure of free atoms. Despite the atom-solid interaction, the 6s6p $^3P_0$ level is found to remain metastable with its lifetimes determined to be in the range of ten to hundreds of seconds. The atomic population can be almost completely transferred between the ground level and the metastable level via optical excitation and spontaneous decay. The dynamics of this process is examined, and is used to explicitly demonstrate that the transition broadening mechanism is homogeneous.

An inert solid, formed by molecular hydrogen or a noble gas element at a cryogenic temperature, offers an interesting and useful medium for atomic and molecular studies. On the one hand, the host medium binds to the guest atoms and molecules, providing confinement and long storage time. On the other hand, due to its chemical inertness, the medium only weakly disturbs the internal structure of the guest atoms and molecules, allowing them to qualitatively retain the structure and transition properties of those free atoms and molecules. This technique, commonly called matrix isolation, has been applied to observe the spectra of a large number of atomic [1–5] and molecular species [6–8]. It is particularly useful for studying chemically reactive species that are difficult to maintain in the gas phase [9, 10].

The atomic spectra obtained with the matrix isolation technique showed that atomic transitions in a noble gas solid are typically blue shifted and broadened to widths several orders of magnitude larger than the natural transition ones of atoms in vacuum [1-5]. Consequently, matrix isolation is not suitable for precision transition frequency measurements. Instead, the technique has potential for applications including studying rare isotopes [11], catching and detecting rare decay products [12], and testing fundamental symmetries [13, 14]. The prospect of measuring the electron's electric dipole moment in a matrix has motivated a series of studies of alkali atoms embedded in solid helium, in which a long electronic spin relaxation time was observed [15], the optical detection of nonradiating atoms was realized [16], and lifetimes of the short-lived excited levels were measured [17].

In this paper, we report the findings of a study of the optical excitation and decay dynamics of neutral ytterbium (Yb) atoms embedded in a solid neon (sNe) matrix at 4 K. In contrast to the simpler alkali atoms that were studied in prior work, the ytterbium atom has two outer electrons that form both spin-singlet and -triplet states (Fig. 1). The lowest-lying triplet levels, 6s6p $^3P_{0,1,2}$, offer additional opportunities to study atom-medium interactions. Moreover, with a nuclear spin of ½, $^{171}$Yb (isotopic

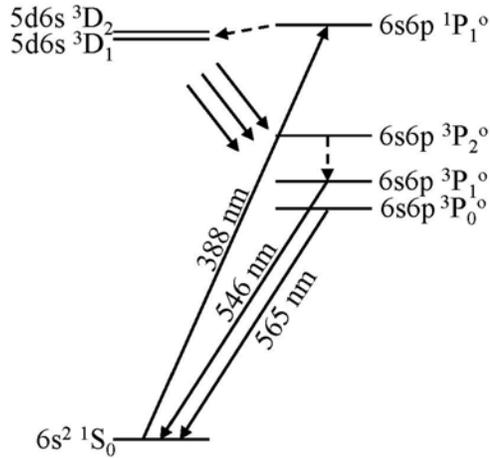

FIG. 1. Low-lying atomic levels and transition wavelengths of Yb in solid neon. Observed (inferred) transitions are shown in solid (dashed) arrows.

abundance = 14%) is an ideal candidate to study nuclear magnetic resonance and perhaps to search for the nuclear Schiff moment [14, 18] in a solid matrix. Optical absorption spectra of Yb embedded in argon, krypton, and xenon matrices have previously been reported [5]. We chose solid neon for the medium because it is less polarizable than heavier noble gas elements, and it is more technically accessible than solid helium. While helium only solidifies under high pressure (> 25 bars), neon readily forms a solid with a face-centered-cubic crystal structure under zero pressure at 4 K [19].

The optical absorption spectrum recorded in this work indicates that neutral ytterbium atoms embedded in solid neon qualitatively retain the structure of free atoms. Both the singlet-to-singlet allowed transitions and the singlet-to-triplet intercombination transitions have been identified. The atomic transition resonances are found to be shifted by and the widths broadened to a few hundred wavenumbers (in the unit of $cm^{-1}$). In addition to the spectrum, the excitation and decay processes of the embedded atoms are studied in this work. Notably, the $6s6p\ ^3P_0$ level is found to remain metastable. Its decay to the ground level follows a multi-exponential function with three time constants: 17±2 s, 67±10 s and 315±50 s. The multitude of decay lifetimes are attributed to the difference between odd and even isotopes and to different types of lattice sites where Yb atoms reside. The atomic population can be almost completely transferred between the ground level $6s^2\,^1S_0$ and the long-lived metastable level $6s6p\,^3P_0$ via optical excitation and spontaneous decay. The dynamics of this process is examined and is used to demonstrate explicitly that the transition broadening mechanism is homogeneous. The rate of population transfer from the singlet ground level to the triplet metastable level is found to be enhanced over that of free atoms by seven orders of magnitude, a phenomenon that can possibly be explained by Stark-induced mixing due to a strong local electric field associated with the matrix.

The Yb-sNe sample is grown on a 2.5 cm diameter $CaF_2$ substrate attached to, and on the vacuum side of, a 4 K cold plate underneath a standard liquid helium cryostat. $CaF_2$ is chosen for its optical transparency in the visible-UV range and for its excellent thermal conductivity at low temperature. Inside the vacuum system, neon atoms are sprayed onto the substrate from a nearby feed tube. The growth of the neon thin film is monitored by the modulated transmission of a 655 nm laser beam. There is an upper limit to the growth rate beyond which the transparency of the neon sample drops significantly. Under ideal conditions, a typical transparent solid neon sample of 2.5 cm diameter and 30 μm thickness is grown in three hours. During the growth of the neon solid, an Yb atomic beam is directed at the substrate from a thermal oven located 0.5 m away. The Yb atomic beam flux in vacuum is measured by the laser induced fluorescence and controlled by adjusting the oven temperature around 300 °C in order to achieve an Yb/Ne atomic ratio of about $1\times10^{-6}$. Under these conditions, the average spacing between adjacent Yb atoms is 44 nm and the Yb areal density is $(1.4\pm0.7)\times10^{14}$ cm$^{-2}$.

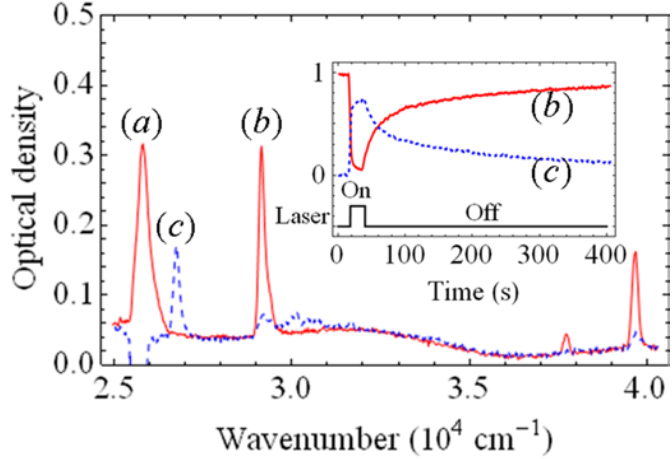

FIG. 2 (color online). Absorption spectra of Yb in solid neon. Red solid line: excitation laser off. Four absorption peaks are assigned to four transitions all originating from the ground level. Blue dashed line: excitation laser on. Upon laser excitation, the original four peaks all disappear; meanwhile, a new peak (*c*) corresponding to 6s6p $^3P_0$ – 6p$^2$ $^3P_1$ appears. The apparent negative optical density below peak (*a*) is due to scattered light of the excitation laser. Inset: Areas of absorption peaks (*b*, red solid line) and (*c*, blue dashed line) vs. time as the excitation laser is switched on and off.

TABLE I. Observed transitions of Yb in solid neon. Here $\sigma$ represents transition wavenumbers in sNe, $\Delta\sigma$ the wavenumber shifts from the in-vacuum values, and $w$ the transition widths. The wavenumber values are in cm$^{-1}$ with an uncertainty of $\pm 40$ cm$^{-1}$.

| Peak | Transition | $\sigma$ | $\Delta\sigma$ | $w$ |
|------|------------|----------|----------------|-----|
| (*a*) | 6s$^2$ $^1S_0$–6s6p $^1P_1$ | 25,790 | + 720 | 350 |
| (*b*) | 6s$^2$ $^1S_0$–4f$^{13}$5d6s$^2$ | 29,140 | + 280 | 140 |
| (*c*) | 6s6p $^3P_0$–6p$^2$ $^3P_1$ | 26,740 | + 220 | 170 |
| (*d*) | 6s$^2$ $^1S_0$–6s6p $^3P_1$ | 18,320 | + 420 | 180 |
| (*e*) | 6s$^2$ $^1S_0$–6s6p $^3P_0$ | 17,700 | + 410 | 250 |

Optical absorption spectroscopy is performed on the sample with a deuterium and tungsten halogen lamp as the white-light source and an optical spectrometer with a wavelength resolution of 1.5 nm. The spectrum (Fig. 2 red solid line) covering the range of 250 – 400 nm shows four distinct absorption peaks identified as atomic transitions of Yb, all originating from the ground level $6s^2\ {}^1S_0$. Among them, two peaks (Fig. 2, peak *a*, *b*) are studied in depth with their properties listed in Table I. For the other two peaks near 40,000 cm$^{-1}$, the weaker transition is identified as $6s^2\ {}^1S_0 - 4f^{13}5d^26s$ and the stronger one as $6s^2\ {}^1S_0 - 6s7p\ {}^1P_1$. The absorption signal on the $6s^2\ {}^1S_0 - 6s6p\ {}^1P_1$ transition (peak *a*) is used to independently determine the areal Yb density. Under the assumption that the oscillator strength of this E1 allowed transition is unmodified by the sNe matrix, the Yb density is calculated based on the absorption signal to be $(0.9\pm0.1)\times10^{14}$ cm$^{-2}$. This is in good agreement with the value $(1.4\pm0.7)\times10^{14}$ cm$^{-2}$ determined from the Yb atomic beam flux and deposition time. This agreement validates the assumption that the oscillator strength and the spontaneous decay lifetime associated with this E1 allowed transition are not significantly changed in the medium. Previous measurements of excited state lifetimes for alkali atoms in solid helium were found to be within 9% of the in-vacuum value [17].

While monitoring the white-light absorption spectrum of the sample, a narrow bandwidth (~ 1 MHz) laser beam at 388 nm, produced by frequency doubling a single-mode Ti:Sapphire ring laser, is used to resonantly excite the $6s^2\ {}^1S_0 - 6s6p\ {}^1P_1$ transition (peak *a*). For a laser intensity of only 2 mW/cm$^2$, all four absorption peaks disappear almost completely in a few seconds while a new absorption peak (*c*) emerges (Fig. 2 blue dashed line). Once the excitation laser beam is blocked, the original four peaks recover and the new peak disappears in a few minutes. The dynamics of peaks (*b*) and (*c*) are shown in the inset of Fig. 2. The opposite and matching time evolution of these two peaks strongly suggests that the atomic population is being transferred between the ground level and a metastable level. Based on a study of the fluorescence dynamics, described in the next section, the metastable level is identified as 6s6p $^3P_0$, and peak (*c*) as transition 6s6p $^3P_0$ – 6p$^2$ $^3P_1$ (Table I).

For a closer examination of the metastable levels, we use the spectrometer to monitor the fluorescence from the 6s6p $^3P_{0,1,2}$ levels following the switch-on and -off of the 388 nm excitation laser. The white-light lamp is kept off in this part of the experiment. The spectrum presented in the upper panel of Fig. 3 is integrated over 0.4 s immediately following the switch-on of the excitation laser. Here the most prominent fluorescence peak (*d*) is identified as 6s$^2$ $^1S_0$ – 6s6p $^3P_1$ (Table I), the only E1 allowed transition in this $^1S_0$ – $^3P_{0,1,2}$ intercombination group. The spectra given in the lower panel are integrated over 1 s after the excitation laser has already been switched off for 5 s (solid line) and 20 s (dashed line). The previously dominant peak (*d*) disappears upon the switch-off of the excitation laser, allowing the spectrum to reveal two much weaker, but long-lived fluorescence peaks from the decay of metastable states. Assuming similar wavenumber shifts of peaks for the $^3P_{0,1,2}$ group, peak (*e*) is identified as 6s$^2$ $^1S_0$ – 6s6p $^3P_0$ (Table I), a transition known to be forbidden in vacuum. Apparently 6s6p $^3P_0$ remains metastable despite the atom-solid interaction. The decay of the fluorescence signal on peak (*e*) after the excitation laser has been switched off is recorded with both the spectrometer and, for improved sensitivity, a photodiode behind an interference filter (Fig. 4). A fit to the decay curve reveals three lifetimes: 17±2 s, 67±10 s and 315±50 s. Moreover, the decay of the unidentified peak at 18,250 cm$^{-1}$ (Fig. 3 lower panel) is best described by a single exponential lifetime of 8±2 s.

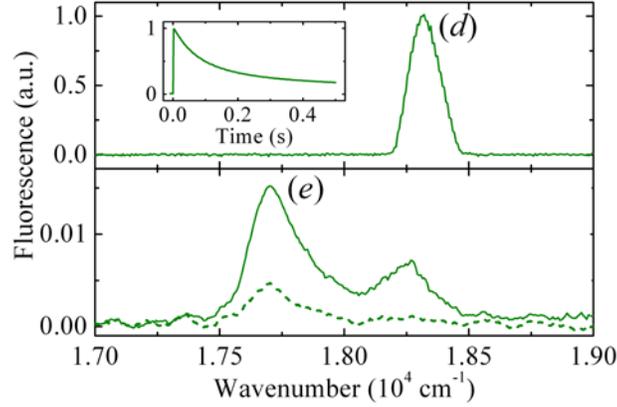

FIG. 3. (color online) Fluorescence spectra of Yb in solid neon. Upper panel: spectrum integrated over 0.4 s immediately following the switch-on of the excitation laser. Peak (*d*) is identified as $6s^2\ ^1S_0 - 6s6p\ ^3P_1$. Inset: The area of fluorescence peak (*d*) vs. time following the switch-on of the excitation laser. The 0.2±0.1 s decay tail indicates the time scale for the transfer of atomic population from the ground level to the metastable level. Lower panel: spectra integrated over 1 s after the excitation laser has been swtiched off for 5 s (solid line) and 20 s (dashed line). Peak (*e*) is identified as $6s^2\ ^1S_0 - 6s6p\ ^3P_0$.

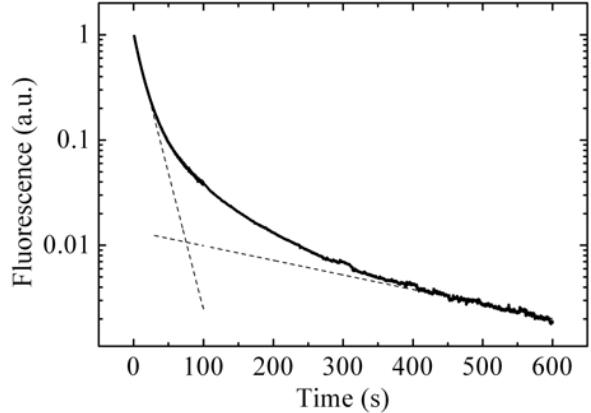

FIG. 4. (color online) The area of fluorescence peak (*e*) vs. time following the switch-off of the excitation laser. A fit reveals three decay time constants: 17±2 s, 67±10 s and 315±50 s. The shortest one is attributed to the mean lifetime of $6s6p\ ^3P_0$ in odd isotopes; the two longer ones are attributed to the lifetimes in even isotopes, possibly from atoms embedded in two different types of lattice sites. The dashed lines represent the 17 s and 315 s exponential decays.

For an Yb atom in vacuum, 6s6p $^3P_0$ is metastable since its decay to the ground level is strictly forbidden. The metastability is partially quenched due to hyperfine interaction in odd isotopes ($^{171}$Yb and $^{173}$Yb, total abundance = 30%), which leads to a lifetime calculated to be around 25 s [20, 21]. An Yb atom in sNe is strongly affected by both the large E-field and E-field gradient present in the lattice, which may induce decay of the metastable 6s6p $^3P_0$ level. The measured time constant, 17±2 s, is attributed to the mean lifetime of odd Yb isotopes embedded in the sNe matrix. The long time constants, 67±10 s and 315±50 s, are attributed to the lifetimes of even Yb isotopes in sNe, possibly from atoms located on two different types of lattice sites.

With the metastable level identified as 6s6p $^3P_0$, the previously described appearance and disappearance of absorption peaks can be understood better. Once the 388 nm excitation laser beam is switched on, it resonantly excites the Yb atoms from the ground level 6s$^2$ $^1S_0$ to 6s6p $^1P_1$, from which the atoms can leak to 5d6s $^3D_{1,2}$, then to 6s6p $^3P_{0,1,2}$, and eventually settle down in the metastable 6s6p $^3P_0$ level. As the population is transferred to the metastable level, the four original absorption peaks disappear; the new absorption peak (*c*) corresponding to a transition from the metastable level (6s6p $^3P_0$ – 6p$^2$ $^3P_1$) appears. As supporting evidence, we observed the fluorescence at 1.3 μm wavelength corresponding to 5d6s $^3D_{1,2}$ – 6s6p $^3P_{0,1,2}$ transitions upon excitation at 388 nm. Moreover, we shined 374 nm light from an LED onto the sample to excite peak (*c*), and observed quenching of the metastable population: the four original absorption peaks re-appear and peak (*c*) disappears completely in only a few seconds. This is a useful tool in the aforementioned study as the atomic population can be quickly reset back to the ground level at the end of each measurement cycle.

In contrast to the metastable 6s6p $^3P_0$ level, the 6s6p $^3P_1$ level decays to the ground level via an E1 transition with a lifetime of 0.9 μs in vacuum [22]. Due to this short lifetime, an equilibrium of population exchange is quickly established between 6s6p $^3P_1$ and the ground level 6s$^2$ $^1S_0$. Therefore, the depletion of the ground level population should also be reflected in the 6s6p $^3P_1$ fluorescence power. This is indeed observed in the evolution of the fluorescence peak (*d*) (Fig. 3 inset). The initial rapid build-up following the switch-on of the 388 nm excitation laser is due to the short lifetime of 6s6p $^3P_1$. The following decay with a 0.2±0.1 s tail manifests the gradual transfer of the population from the ground level 6s$^2$ $^1S_0$ to the metastable level 6s6p $^3P_0$.

The study on the dynamics of the atomic excitation and population transfer led us to conclude that the broadening mechanism observed in both absorption and fluorescence spectra is homogeneous. The bandwidth of the excitation laser is 1 MHz; the natural linewidth of the 6s$^2$ $^1S_0$ – 6s6p $^1P_1$ transition at 388 nm is 30 MHz. In comparison, the observed absorption linewidth is 350 cm$^{-1}$, or $4\times10^5$ times the natural linewidth. The broadening must be dominated by a homogeneous mechanism for the transfer of the entire population between the ground level and the metastable level to be possible. We hypothesize that the broadening is due to strong dephasing interactions between the Yb atoms and the sNe matrix. Also referred to as phonon broadening, this mechanism is an important source of homogenous broadening in solid state lasers [23]. Collisions between the Yb atoms and the phonons in the sNe matrix dephase but do not quench the excited state. This explains how the linewidth can be very broad without significantly reducing the lifetime.

The observed linewidth of the excitation transition 6s$^2$ $^1S_0$ – 6s6p $^1P_1$ is broadened by a factor of $4\times10^5$. This should translate into a reduced excitation rate of the resonant laser by the same factor. At 2 mW/cm$^2$, Yb atoms in sNe have a calculated excitation rate of 10 s$^{-1}$. Meanwhile, measurements indicate that the atomic population is transferred from the ground level 6s$^2$ $^1S_0$ to the metastable level 6s6p $^3P_0$ in an average time of 0.2 s or at a rate of 5 s$^{-1}$. This suggests an abnormally high singlet-to-triplet transition branching ratio of approximately 50%. In vacuum, 6s6p $^1P_1$ decays to 5d6s $^3D_{1,2}$ with a branching ratio of $7\times10^{-8}$. For Yb in sNe, the singlet-to-triplet decay branching ratio appears to be enhanced by seven orders of magnitude, a phenomenon that can possibly be explained by Stark-induced mixing due to a strong E-field and E-field gradient present in the matrix. However, this naïve model generates contradictory results: the electric field needed to explain the enhanced decay branching ratio would also significantly quench the metastable level 6s6p $^3P_0$ due to second order effects. A sophisticated theoretical model taking into account the Yb-sNe interaction is needed to quantitatively explain this and other observations reported in this paper.

We would like to thank T. Oka for the gift of the cryostat and for many helpful discussions throughout the project. This work is supported by Department of Energy, Office of Nuclear Physics, under contract DEAC02-06CH11357. S.-M. Hu acknowledges the support from NKBRSF (2007CB815203).